\documentclass[aps,prb,numerical,superscriptaddress,showpacs,showkeys,reprint,byrevtex]{revtex4-1} 
\usepackage[utf8x]{inputenc}
\usepackage{ucs}
\usepackage{epsfig}
\usepackage{amsmath}
\usepackage{caption}
\usepackage{subcaption}
\usepackage{amsfonts}

\usepackage{amssymb}
\usepackage{graphicx}
\usepackage{dcolumn}
\usepackage{bm}
\usepackage{latexsym}
\usepackage{hyperref}
\usepackage{color}
\usepackage{bm}
\usepackage{amsmath}
\usepackage{amsfonts}
\usepackage{amssymb}
\usepackage{graphicx}
\usepackage{setspace}

\def\bra#1{\mathinner{\langle{#1}|}}
\def\ket#1{\mathinner{|{#1}\rangle}}

\begin{document}
\title{Bound state and persistent currents in the presence of torsion and Rashba spin-orbit coupling}
\author{Debabrata Sinha}
\affiliation{TIFR Centre for Interdisciplinary Sciences, Hyderabad 500075, India}
\email{debabratas@tifrh.res.in}

\begin{abstract}
We study a model of an electron on a cylindrical surface, which is coupled to the torsion field due to a dislocation along the axis of the cylinder. We discuss the effect of this torsion field on the energy spectrum of the electrons and analytically calculate persistent currents in the presence of Rashba spin-orbit coupling. We also analyze bound state energy spectra in presence of Rashba spin-orbit interaction. Our results show that the presence of the torsional field due to the dislocation significantly modifies the energy spectrum of the system. The dislocation induced persistent spin current in this system is calculated, and we find a correspondence between the dislocation mediated spin current and the azimuthal spin current. 
\end{abstract} 

\maketitle

The effect of spin-orbit(SO) coupling on electrons has been studied in several contexts such as spintronics, magnetoconductance, mesoscopic ring and quantum spin hall effect \cite{Wolf},\cite{Zutic},\cite{Meso-ring},\cite{PRB66},\cite{Pers},\cite{PRL96}. SO couples the spin degree of freedom of an electron to its orbital motion, thereby giving rise to a useful way to manipulate and control the electron spin by an external field. In semiconductor based nanostructures, the Rashba SO coupling arises due to the structural asymmetry of the host crystal\cite{Rashba}. The quantum confinement of electrons in a ring using SO coupling helps us understand many important quantum effects of electrons. In recent times, SO coupling has been extensively studied in a quantum ring for understanding various quantum interference phenomena such as the existence of persistent currents, Aharonov-Bohm effect, spin relaxation, etc \cite{Meso-ring},\cite{Pers},\cite{Spin-relax}. Such effects have potential application in the fields of quantum computation and quantum information. 

Although, most theoretical studies are based on the one dimensional quantum ring, the two dimensional model has also been investigated\cite{Bulaev},\cite{Tan}. The behaviour of a quantum particle confined within a curved surface is significantly different from that on a flat surface. Due to the confinement, the excitation energy normal to the surface is considerable higher than in the tangential direction. One can write down the effective Hamiltonian that involves the curvature-induced scalar potential\cite{Maraner},\cite{Costa}. The effect of curvature has been studied in the context of condensed matter\cite{PRB64},\cite{Entin},\cite{PRB72},\cite{PRL100},\cite{PRB79}, owing to the recent progress in nanotechnology that has made possible the production of curved two dimensional layers of several geometries and shapes\cite{Ono},\cite{Shea},\cite{Schmidt}. In addition to surface curvature, geometric torsion is another important parameter that has influence on the low dimensional quantum mechanical properties of a particle\cite{Torsion1},\cite{Torsion2},\cite{Berry-Phase}.
 
The correspondence between the theory of defects in solids and three dimensional gravity is well established\cite{Katanaev}. The relevance of torsion and the symmetry of the stress tensor in a solid was noticed by Cartan \cite{Cartan}. Since then the Riemann-Cartan geometry in a solid has been of growing interest for condensed matter physicists. In the continuum limit, the solid is described by the Riemann-Cartan manifold where the Burgers vector of a dislocation is associated with the torsion present, and the Frank angle of disclination corresponds to the curvature. When an object traverses a loop in real space, it is rotated if there is non zero curvature and translated if there is non zero torsion. The phase shift due to torsion in the presence of dislocation may give rise  to a novel class of persistent currents that can flow along the dislocation line. Quantitative discussion on the torsional effect on the energy spectrum and persistent current is important from both theoretical and practical points of view.

In this contribution, we examine the modification of the energy spectrum of a cylindrical wire in the presence of Rashba spin-orbit coupling and the torsional field of a single topological defect. We specifically consider a screw dislocation and confine the electrons to a cylindrical surface enclosing it. The non-trivial topology created by the defect breaks the degeneracies in the energy spectrum which would otherwise be present. We calculate the torsion induced persistent current and find its correspondence with the surface current.
 
\begin{figure}
\begin{center}
\rotatebox{0}{\includegraphics[width=0.8in]{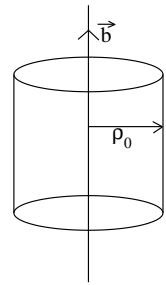}} 
\end{center}
\caption{Electrons are confined on cylindrical surface of radius $\rho_{0}$, enclosing a screw dislocation along its axis. The dislocation is characterized by a Burgers vector $\vec{b}$.}
\label{model}
\end{figure}

We now describe our model in detail. We consider a single screw dislocation, characterized by a Burgers vector {\bf b}. This produces a strain field which, to first order, is $u_{ij}=\partial_iu_j+\partial_ju_i$, where $u_i$ is the displacement of a lattice point from its position in the defect-free medium. The metric of the deformed system is given by $g_{ij}=\delta_{ij}+u_{ij}$. The Hamiltonian for the propagation of a free-electron in this space is $p_ig^{ij}p_j/2m$ where ${\bf p}$ is the momentum operator, $m$ the mass of an electron and $g^{ij}$ the inverse of $g_{ij}$.

The description of the spin-orbit coupling in the presence of defects is more complicated since it is sensitive to the local torsion and the curvature of the space. However we can define a locally flat, $x^{a}(a=1,2,...,d)$ frame to our manifold at all points excluding the dislocation. The components of the matrix ${e^{a}}_{i}$, for the transformation from the coordinate $x^{i}$, is defined over the whole manifold where ${e^{a}}_{i}={\partial x^{a} \over \partial x^{i}}$. The components ${e^{a}}_{i}$ are the tetrad and contains information about the dislocation via the line-integral: $\int_{c} {e^{a}}_{i} dx^{i}=\int_{c} de^{a}=-b^{a}$, where $b$ is the magnitude of the Burgers vector enclosed within the curve $c$. In a lattice system, ${e^{a}}_{i}$ defined on each lattice site encodes the local bond stretching and the local orbital orientation of the system. The inverse of the tetrad, ${e_{a}}^{i}$, is defined through the orthogonality relation ${e_{a}}^{i}{e^{a}}_{j}={\delta^{i}}_{j}$. The tetrad satisfy the completeness relation: ${e^{a}}_{i}{e_{b}}^{i}={\delta^{a}}_{b}$ and are linked to the metric, $g_{i j}$,through $g_{i j}=\eta_{a b}{e^{a}}_{i}{e^{b}}_{j}$. An electron with SO coupling propagating in a lattice will be sensitive to the local orbital orientation. In this case, the electron momentum is rescaled as $p_{i}\rightarrow p_{i}{e^{i}}_{a}$\cite{Taylor}. 

 The screw dislocation can be understood in terms of the Volterra process, where one starts with a perfect crystal and a suitable plane inside the crystal is chosen for an incision. The crystal on one side of the plane is then displaced by a lattice vector (Burgers vector), and additional atoms are inserted or removed. In this way, the crystalline order is restored everywhere except the line of incision. The distance $(ds)$ between two points in the medium with infinite long screw dislocation can be written as \cite{EPL45}
\begin{eqnarray}
ds^{2}=g_{ij}dx^{i}dx^{j}=d\rho^{2}+\rho^{2}d\phi^{2}+(dz+\beta d\phi)^{2}
\label{dis}
\end{eqnarray}
with $\rho > 0$, $0\leq \phi \leq 2\pi$ and $-\infty \leq z \leq \infty$. The effect of the Burgers vector enters as $\beta=\frac{b}{2\pi}$. Correspondingly, from eq(\ref{dis}), we get the inverse of metric to be  given by
\begin{eqnarray}
g^{ij}=
\begin{pmatrix}
1& 0 & 0\\
0& 1 & -\frac{\beta}{\rho}\\
0&-\frac{\beta}{\rho}& \frac{\beta^{2}+\rho^{2}}{\rho^{2}}
\end{pmatrix}
\label{gij}
\end{eqnarray}
and the frame field $e^{i}_{a}$ is
\begin{eqnarray}
e^{i}_{a}=
\begin{pmatrix}
\cos \phi & \sin \phi & 0\\
-\frac{\sin \phi}{\rho} & \frac{\cos \phi}{\rho} & 0\\
\frac{\beta}{\rho}\sin \phi & -\frac{\beta}{\rho}\cos \phi & 1
\end{pmatrix}
\label{triad}
\end{eqnarray}
We begin with the 3D Hamiltonian of an electron with Rashba spin-orbit interaction in the absence of dislocations as
\begin{eqnarray}
H^{\pm}_{3D}=\frac{p^{2}}{2m}\pm \omega(p_{x}\sigma_{y}-p_{y}\sigma_{x})
\label{rashba}
\end{eqnarray} 
where, $\sigma_{i}$ is a Pauli matrix and $\omega$ characterizes the strength of Rashba SO interaction. The $3D$ Hamiltonian in eq(\ref{rashba}) can be rederived from the $3D$ relativistic Dirac equation. In the $3D$ case, the relativistic Dirac Hamiltonian is \cite{Wu}
\begin{eqnarray}
H^{3D}_{Dirac}=\frac{\sqrt{l_{so}}\omega}{\sqrt{2}}
\begin{pmatrix}
0 & \vec{\sigma}.\vec{p}+\frac{i\hbar}{l_{so}}\sigma_{z}\\
\vec{\sigma}.\vec{p}-\frac{i\hbar}{l_{so}}\sigma_{z} & 0
\end{pmatrix}
\label{dirac}
\end{eqnarray}
We define the length scale for spin-orbit coupling as $l_{so}=\frac{\hbar}{m\omega}$. The square of eq(\ref{dirac}) exhibits block-diagonal form
\begin{eqnarray}
\frac{H^{2}_{3D}}{\hbar \omega}-\frac{1}{2}m\omega^{2}=
\begin{pmatrix}
H^{-}_{3D} & 0\\
0 & H^{+}_{3D}
\end{pmatrix}
\end{eqnarray}
Along similar lines, in the presence of a dislocation, the relativistic Dirac Hamiltonian takes the form
\begin{eqnarray}
H^{3D}_{Dirac}=\frac{\sqrt{l_{so}}\omega}{\sqrt{2}}
\begin{pmatrix}
0 & (p_{i}e^{i}_{a}\sigma^{a}+\frac{i\hbar}{l_{so}}\sigma_{z})\\
(p_{i}e^{i}_{a}\sigma^{a}-\frac{i\hbar}{l_{so}}\sigma_{z}) & 0
\end{pmatrix}\nonumber\\
\end{eqnarray}
This results in the Hamiltonian $H^{-}_{3D}$ to be defined as follows
\begin{eqnarray}
H^{-}_{3D}=\frac{1}{2m}p_{i}g^{ij}p_{j}+i\omega p_{i}e^{i}_{a}[\sigma_{z},\sigma_{a}]
\label{SO-hamil}
\end{eqnarray}
describes the motion of an electron in the presence of a dislocation and Rashba SO interaction. The metric $g^{ij}$ and frame field $e^{i}{a}$ for screw dislocation have been specified earlier in eq(\ref{gij}) and eq(\ref{triad}) respectively. The Hamiltonian in eq(\ref{SO-hamil}) possesses time reversal symmetry, and translational symmetry along the dislocation axis i.e. the $z$-axis. We label the states in the $z$-direction by $k$, the crystal momentum in this direction. 

We introduce a circular harmonic potential $V(\rho)=\frac{1}{2}m\omega^{2}(\rho-\rho_{0})^{2}$ to confine the electrons on a cylindrical surface of constant radius $r=\rho_{0}$, where $\rho_{0}$ is much larger than the dislocation core radius and the electron's wavelength. We suppose that the electrons are confined to the lowest radial band $\ket{R_{0}}$ of the Hamiltonian
\begin{eqnarray}
H_{3D}(\rho)=-\frac{\hbar^{2}}{2m}(\partial^{2}_{\rho}+\frac{1}{\rho}\partial_{\rho})+V(\rho)
\end{eqnarray}
where the potential $V(\rho)$ is assumed to be steep. To construct a Hamiltonian which defines the electrons on the cylindrical surface, we project the Hamiltonian on the lowest radial band:
\begin{eqnarray}
\mathcal{H}(\phi,z)=\bra {R_{0}}H^{-}_{3D}-K.E.\ket{R_{0}}
\label{correct-hamil}
\end{eqnarray}
where K.E. is the kinetic energy associated with the radial motion of the electron. By explicit calculation, it can be shown that the matrix elements $\bra {R_{0}} \rho^{-1} \ket {R_{0}}=\rho^{-1}_{0}$ and $\bra {R_{0}} \partial\rho \ket {R_{0}}=-(2\rho_{0})^{-1}$ where $\ket {R_{0}}$ is the eigenstate of lowest radial band \cite{PRB66}. Using the above identities, the surface Hamiltonian eq(\ref{correct-hamil}) reduces to
\begin{eqnarray}
\mathcal{H}(\phi,z)&=&-\frac{\hbar^{2}}{2m}\partial_{z}^{2}+\frac{\hbar^{2}}{2m\rho^{2}_{0}}[(i\partial_{\phi}-i\beta \partial_{z}+\frac{m\omega \rho_{0}}{\hbar}\sigma_{\rho})^{2}\nonumber\\&&-(\frac{m\omega \rho_{0}}{\hbar})^{2}]
\label{shamiltonian}
\end{eqnarray}
with $\sigma_{\rho}=\cos \phi \sigma_{x}+\sin \phi \sigma_{y}$ and $\sigma_{\phi}=-\sin \phi \sigma_{x}+\cos \phi \sigma_{y}$.

In order to solve the eigenvalue equation $\mathcal{H}(\phi,z)\Psi(\phi,z)=E\Psi(\phi,z)$, we choose the general form of wavefunction $\Psi(\phi,z)=e^{i\lambda n\phi}e^{iskz}\xi_{\sigma}$, where $n$ is the principal quantum number and $\lambda,\sigma, s=\pm 1$. $\lambda=\pm$ corresponds to clockwise and anticlockwise propagating waves. $\sigma=\pm$ corresponds to $\sigma_z=\pm$. $s=\pm$ corresponds to waves propagating along $z$ and $-z$ direction. $\xi_{\sigma}$ is the spinor wavefunctions.
The eigenvalues of eq(\ref{shamiltonian}) are
\begin{eqnarray}
E^{n,\lambda}_{s,\sigma}&=&\frac{\hbar^{2}}{2m\rho^{2}}[(n-s\lambda\beta k-\frac{\lambda\sigma}{2}\sqrt{1+4\frac{m^{2}\omega^{2}\rho^{2}_{0}}{\hbar^{2}}})^{2}+k^{2}\rho^{2}_{0}\nonumber\\&&-(\frac{m\omega\rho_{0}}{\hbar})^{2}]
\label{eng-spc}
\end{eqnarray}
with $n$ is an half odd integer. 
\begin{table}[ht]
\caption{Energy in unit of $\frac{\hbar^{2}}{2m\rho^{2}}$.In the absence of SOC and dislocation i.e. $\omega=0$ and $\beta=0$, energy is $E^{n,\lambda}_{s,\sigma}=\frac{\hbar^{2}}{2m\rho^{2}}(n-\lambda \sigma/2)=\frac{\hbar^{2}N^{2}}{2m\rho^{2}}$. In presence of SOC and dislocation, the energy degeneracies are broken. We have shown here only for first few energy levels. $E1$ denotes the first four degenerate energy levels, $E2$ the second four degenerate levels, $E3$ the next two degenerate levels and $E4$ the last two degenerate levels. Full energy spectrum can be calculated from eq(\ref{eng-spc}).}
\centering
\begin{tabular}{c| c c c c c c|}
\hline\hline
$E(\omega,\beta=0)$ & $n$ & $\lambda$ & $\sigma$ & $s$ & $N$ & $E-k^{2}\rho^{2}_{0}+\Omega^{2}(\omega,\beta \neq 0)$\\
\hline 
0 & -1/2 & + & - &- & 0 & $(1/2-1/2\Omega_{1}-\beta k)^{2}$\\
0 & -1/2 & - & + &+ & 0 & $(1/2-1/2\Omega_{1}-\beta k)^{2}$\\
0 &  1/2 & + & + &+ & 0 & $(1/2-1/2\Omega_{1}-\beta k)^{2}$\\
0 &  1/2 & - & - &- & 0 & $(1/2-1/2\Omega_{1}-\beta k)^{2}$\\
\hline
0 & -1/2 & + & - &+ & 0 & $(1/2-1/2\Omega_{1}+\beta k)^{2}$\\
0 & -1/2 & - & + &- & 0 & $(1/2-1/2\Omega_{1}+\beta k)^{2}$\\ 
0 &  1/2 & + & + &- & 0 & $(1/2-1/2\Omega_{1}+\beta k)^{2}$\\
0 &  1/2 & - & - &+ & 0 & $(1/2-1/2\Omega_{1}+\beta k)^{2}$\\
\hline
1 &  1/2 & + & - &+ & 1 & $(1/2+1/2\Omega_{1}-\beta k)^{2}$\\
1 &  1/2 & - & + &- & 1 & $(1/2+1/2\Omega_{1}-\beta k)^{2}$\\
\hline
1 &  1/2 & - & + &+ & 1 & $(1/2+1/2\Omega_{1}+\beta k)^{2}$\\
1 &  1/2 & + & - &- & 1 & $(1/2+1/2\Omega_{1}+\beta k)^{2}$\\
\hline
\end{tabular}
$\Omega^{2}=\frac{m^{2}\omega^{2}\rho^{2}_{0}}{\hbar^{2}}$,$\Omega_{1}=\sqrt{1+4\Omega^{2}}$
\label{table}
\end{table}

\begin{figure}
\rotatebox{0}{\includegraphics[width=2.5in]{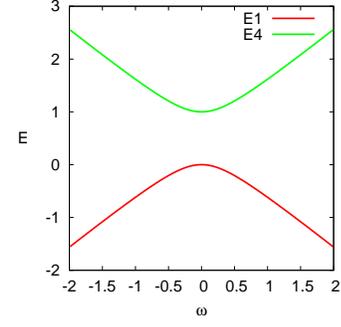}}
\caption{Energy diagram ($E^{n,\lambda}_{s,\sigma}$) as a function of SO interaction strength $\omega$ for $\beta=0, k=0$ and $\rho_{0}=1$. The energy in units of $\frac{\hbar^{2}}{2m\rho^{2}}$ while $\omega$ units of $\frac{\hbar}{m\rho}$. From table (\ref{table}), for $\beta k=0$, all the $N=0$ and $N=1$ energy levels become two different degenerate level.}
\label{fig1}
\end{figure}

\begin{figure}
\rotatebox{0}{\includegraphics[width=2.5in]{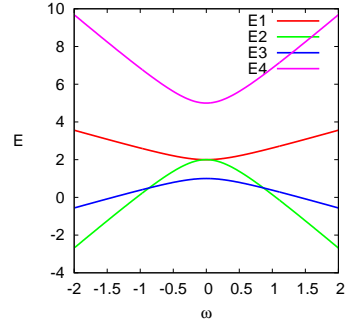}}
\caption{Energy diagram as a function of SO interaction strength $\omega$ for $\beta k=1$ and $\rho_{0}=1$. The energy degeneracies are broken depending on the strength of $\beta k$.}
\label{fig2}
\end{figure}

\begin{figure}
\rotatebox{0}{\includegraphics[width=2.5in]{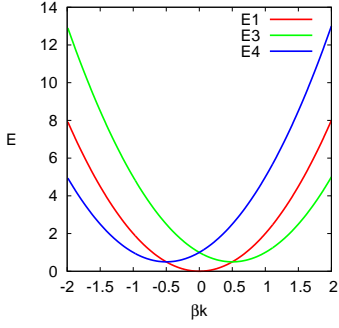}}
\caption{Energy ($E^{n,\lambda}_{s,\sigma}$) spectrum for lowest radial band as a function of $\beta k$ in absence of Rashba term i.e, $\omega=0$.}
\label{fig3}
\end{figure}

\begin{figure}
\rotatebox{0}{\includegraphics[width=2.5in]{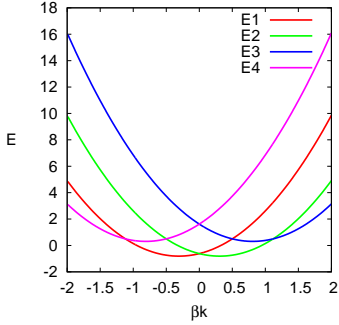}}
\caption{Energy ($E^{n,\lambda}_{s,\sigma}$) diagram as a function of $\beta k$ for Rashba term $\omega=1$. }
\label{fig4}
\end{figure}

In table (\ref{table}), we have calculated the energies for first few energy levels. When $\omega \neq 0$ and $\beta \neq 0$, the energy degeneracies are broken, although maintaining two fold degeneracy of every energy level due to time reversal symmetry. The energy split is larger in the higher energy states. In fig (\ref{fig1}), we have shown the energy spectrum w.r.t $\omega$ for $\beta=0, k=0$ i.e. in the absence of dislocation and $k=0$-th state. From table {\ref{table}}, the $N=0$ and $N=1$ becomes two different degenerate levels i.e. $E1=E2$ and $E3=E4$. In fig(\ref{fig2}) we have shown that, these degenerate energy levels are broken in presence of torsional field. In fig(\ref{fig3}) and fig(\ref{fig4}) we show the effect of a Rashba term on the energy spectrum. In fig(\ref{fig3}), we plot the energy spectrum with $\beta k$ in the absence of Rashba term i.e. $\omega=0$. It is clear that at $\omega=0$, $E1=E2$. This ground state degeneracy is broken for finite value of Rashba coupling which is shown in fig(\ref{fig4}).

The energy eigen-functions of eq(\ref{shamiltonian}), for the eigenvalues in eq(\ref{eng-spc}) are
\begin{eqnarray}
\Psi^{n,+}_{+,+}&=&e^{i n \phi}e^{ikz}
\begin{pmatrix}
\cos(\theta/2)e^{-i\phi/2}\\
\sin(\theta/2)e^{i\phi/2}
\end{pmatrix}\nonumber\\
\Psi^{n,+}_{-,-}&=&e^{i n \phi}e^{-ikz}
\begin{pmatrix}
-\sin(\theta/2)e^{-i\phi/2}\\
\cos(\theta/2)e^{i\phi/2}
\end{pmatrix}\nonumber\\
\Psi^{n,-}_{+,+}&=&e^{-i n \phi}e^{ikz}
\begin{pmatrix}
\cos(\theta/2)e^{-i\phi/2}\\
\sin(\theta/2)e^{i\phi/2}
\end{pmatrix}\nonumber\\
\Psi^{n,-}_{-,-}&=&e^{-i n \phi}e^{-ikz}
\begin{pmatrix}
-\sin(\theta/2)e^{-i\phi/2}\\
\cos(\theta/2)e^{i\phi/2}
\end{pmatrix}
\label{wavefunction}
\end{eqnarray}
with $\tan \theta=2m\omega \rho_{0}/\hbar$.\\
Using the time reversal operator for of spin $1/2$ particle, $\Theta=-iK\sigma_{y}$, it is obvious that $\Psi^{n,+}_{+,+}=\Theta\Psi^{n,-}_{-,-}$ and $\Psi^{n,+}_{-,-}=\Theta\Psi^{n,-}_{+,+}$. Where $K$ is the complex conjugation operator and $\sigma_{y}$ is the Pauli matrix.


We can calculate the persistent spin and charge current by using the ground state wave functions. The charge current is zero since the system has time reversal symmetry. But still there exists a spin current. The existence of the spin current in absence of the dislocation can be explained as follows. We assume that the electron wavelength is much smaller than radius of the cylinder and electron with spin $\sigma$ precesses slowly around the wire. In presence of SO coupling, the electron spin picks up a Berry phase due to its adiabatic motion. The spin Berry phase for the spin up ($\sigma=\uparrow$) electron induces clockwise persistent spin polarized current $I_{1}$. Since the system has time reversal symmetry, the spin down($\sigma=\downarrow$) electron induces anticlockwise persistent spin current $I_{2}$, which is exactly opposite of $I_{1}$ . As a result, the system has net persistent spin currents in the azimuthal direction\cite{PRL98}. Now, in presence of a dislocation, the electron spin acquire an extra phase when it is moving around a dislocation line. This will induce a current along the dislocation line.

From the definition, the charge current density is: $\vec{J}(r)=\psi^{\dagger} e \vec{v} \psi$, where $e$ is the electronic charge and $\vec{v}$ is the velocity operator. The azimuthal and $z$-component velocities are
\begin{eqnarray}
v_{\phi}&=&\frac{i\rho_{0}}{\hbar}[\mathcal{H},\phi]=-\frac{i\hbar}{m\rho_{0}}(\partial_{\phi}-\beta \partial_{z})+\omega \sigma_{\rho}\nonumber\\
v_{z}&=&\frac{i}{\hbar}[\mathcal{H},\phi]=-\frac{i\hbar}{m}\partial_{z}+\frac{i\beta\hbar}{m\rho_{0}}(\partial_{\phi}-\beta \partial_{z})-\frac{\omega \beta}{\rho_{0}} \sigma_{\rho}\nonumber\\
\end{eqnarray}
We are interested in the ground state currents in the system. We project the velocity operator on the $k=0$-th state. The azimuthal and $z$-component velocities for the $k=0$-th state are
\begin{eqnarray}
v^{g}_{\phi}&=&\bra {k=0} v_{\phi} \ket {k=0}=-\frac{i\hbar}{m\rho_{0}}\partial_{\phi}-\omega \sigma_{\rho}\nonumber\\
v^{g}_{z}&=&\bra {k=0} v_{z} \ket {k=0}=\frac{i\hbar\beta}{m\rho^{2}_{0}}\partial_{\phi}+\frac{\omega \beta}{\rho_{0}} \sigma_{\rho}
\end{eqnarray}
Using the wavefunctions in eq(\ref{wavefunction}) for the ground state , it is seen that the azimuthal and z-component charge currents $J_{\phi}=J_{z}=0$.

Now, from the definitions of spin current density $\mathcal{J}^{i}=\frac{1}{2}\psi^{\dagger}\{\hat{\vec{v}},s^{i}\}\psi$, where $s^{i}=\frac{\hbar \sigma^{i}}{2}$, we will calculate the ground state azimuthal and $z$-component spin currents. We project the velocity operator on $k=0$-th state as above and using the wavefunctions in eq(\ref{wavefunction}) for the ground state, the different components of spin current are
\begin{eqnarray}
\mathcal{J}^{z}_{\phi}&=&\frac{\hbar^{2}}{4m\rho_{0}}(\cos \theta -1)\nonumber\\
\mathcal{J}^{x}_{\phi}&=&\frac{\hbar^{2}}{4m\rho_{0}}\sin \theta \cos \phi-\frac{\hbar \omega}{2}\cos \phi\nonumber\\
\mathcal{J}^{x}_{\phi}&=&\frac{\hbar^{2}}{4m\rho_{0}}\sin \theta \sin \phi-\frac{\hbar \omega}{2}\sin \phi
\label{jx}
\end{eqnarray}
The $\phi$ dependent terms in spin currents are due to the precession of the spin around the $z$ axis. The dislocation associated spin currents $\mathcal{J}^{i}_{z}$ are
\begin{eqnarray}
\mathcal{J}^{z}_{z}&=&\frac{\hbar^{2}\beta}{4m\rho_{0}^{2}}(1-\cos \theta)\nonumber\\
\mathcal{J}^{x}_{z}&=&-\frac{\beta \hbar^{2}}{4m\rho_{0}^{2}}\sin \theta \cos \phi+\frac{\omega \beta \hbar}{2\rho_{0}}\cos \phi\nonumber\\
\mathcal{J}^{y}_{z}&=&-\frac{\beta \hbar^{2}}{4m\rho_{0}^{2}}\sin \theta \sin \phi+\frac{\omega \beta \hbar}{2\rho_{0}}\sin \phi
\label{jz}
\end{eqnarray}
The ground state spin currents will be zero if $\cos \theta=1$ i.e. when spin-orbit coupling are absent ($\omega=0$). The z-componet spin current is zero for $\beta=0$ i.e. in absence of dislocation. From eq(\ref{jx}) and eq(\ref{jz}) it is clear that $\mathcal{J}^{i}_{z}=-\frac{\beta}{\rho_{0}}\mathcal{J}^{i}_{\phi}$. Thus, the dislocation induced current in this system are proportional to the azimuthal current. From the eq(\ref{jz}), it is clear that the dislocation mediated spin currents depend only on the Burgers vector of the dislocation. This implies that our result is also true for other kinds of dislocation like edge, mixed etc.

In conclusion, we have discussed the effects of a torsional field on the energy spectrum on a cylindrical surface. We find the correspondence between dislocation mediated spin currents and azimuthal spin currents in the system. Despite the conceptual simplicity, the torsional effect has a significant role in the modification of the surface electronic structure on the 2d curved surface. The most important observation is the presence of a persistent spin current without an accompanying charge current. It is the spin Berry phase due to the SO interaction that causes the finite spin current in the system. The quantum phase shift of an electron in the presence of dislocation induces a spin current along the dislocation line. The dislocation mediated spin current may be used as a probe for persistent spin current.

I am immensely indebted to Siddhartha Lal and Krishnendu Sengupta for many stimulating discussions. Also a word of thanks to Subodh Shenoy and Surajit Sengupta for encouragement and useful discussions. I am grateful to Sharath Jose, Saurish Chakrabarty for helping me with the preparation of the manuscript .

\end{document}